\documentclass[12pt,a4paper,final]{iopart}

\usepackage{iopams}
\usepackage{graphicx}

\usepackage{amssymb}
\usepackage{amscd}	
\usepackage{enumerate}
\usepackage{subfigure}
\usepackage{xcolor}
\usepackage{color}
\usepackage{cases}
\usepackage[margin=1cm]{caption}

\newcommand{\bra}[1]{\mbox{$\left\langle #1 \right|$}}
\newcommand{\ket}[1]{\mbox{$\left| #1 \right\rangle$}}
\newcommand{\braket}[2]{\mbox{$\left\langle #1 | #2 \right\rangle$}}

\newcommand{\eqref}[1]{(\ref{#1})}

\eqnobysec

\begin{document}






\title[Discrete-phase-randomized coherent state source and its application in QKD]{Discrete-phase-randomized coherent state source and its application in quantum key distribution}

\author{Zhu Cao$^{1}$, Zhen Zhang$^{1}$, Hoi-Kwong Lo$^{2}$, Xiongfeng Ma$^{1}$}
\address{$^1$Center for Quantum Information, Institute for Interdisciplinary Information Sciences, Tsinghua University, Beijing, China}
\address{$^2$Center for Quantum Information and Quantum Control, Department of Physics and Department of Electrical \& Computer Engineering, University of Toronto, Toronto,  Ontario, Canada}

\begin{abstract}
Coherent state photon sources are widely used in quantum information processing. In many applications, such as quantum key distribution (QKD), a coherent state is functioned as a mixture of Fock states by assuming its phase is continuously randomized.
In practice, such a crucial assumption is often not satisfied and, therefore, the security of existing QKD experiments is not guaranteed. To bridge this gap, we provide a rigorous security proof of QKD with discrete-phase-randomized coherent state sources. Our results show that the performance of the discrete-phase randomization case is close to its continuous counterpart with only a small number (say, 10) of discrete phases. Comparing to the conventional continuous phase randomization case, where an infinite amount of random bits are required, our result shows that only a small amount (say, 4 bits) of randomness is needed.
\end{abstract}

\vspace{2pc}
\maketitle

\section{Introduction}

In many quantum optics applications, such as QKD~\cite {Bennett:BB84:1984,Ekert:QKD:1991}, linear optics quantum computing~\cite {KLM_01}, bit commitment~\cite{kent2012unconditionally}, coin flipping~\cite {PhysRevA.84.052305}
and blind quantum computing~\cite {PhysRevLett.108.200502}, a perfect single photon source is assumed to be used, which is not feasible with current technology. Instead, a weak laser is widely used to replace the single photon source in practice. A laser can be well described by a coherent state~\cite {PhysRev.131.2766},
\begin{equation} \label{Discrete:CoherentState}
\ket{\alpha}=e^{-\frac{|\alpha|^2}{2}}\sum \frac{\alpha^n}{\sqrt{n!}}\ket{n},
\end{equation}
on which a phase modulation by $\theta\in [0,2\pi)$ implements the operation $\ket{\alpha}$ to
$\ket{\alpha e^{i \theta}}$.
For a coherent state, there is a nonzero probability to get components other than single photons, such as vacuum states and multi photon states. To model this imperfection, a photon number channel model is used~\cite {Lo:Decoy:2005}, which assumes the phase of the coherent state is randomized,
\begin{equation} \label{Discrete:ContinuousPhase}
\frac{1}{2 \pi}\int_{0}^{2\pi}\ket{\alpha e^{i\theta}}\bra{\alpha e^{i\theta}}d\theta =\sum\limits_{n=0}^{\infty}e^{-|\alpha|^2}\frac{|\alpha|^2}{n!}\ket{n}\bra{n}.
\end{equation}
A physical interpretation behind Eq.~\eqref{Discrete:ContinuousPhase} is that when the phase of a coherent state is randomized, it is equivalent to a mixed state of Fock states whose photon number follows a Poisson distribution with a mean of $|\alpha|^2$. In other words, the Fock states are totally decohered from each other with continuous phase randomization.

We remark that phase randomization as specified in Eq.~\eqref{Discrete:ContinuousPhase} is a common assumption in the theoretical models of many quantum information processing protocols.
In practice, as will be discussed further below, the assumption of
continuous phase randomization is often not satisfied
in experiments. Therefore, the security of a protocol
(e.g. the security of a generated key
in QKD) is \emph{not} guaranteed.

To illustrate the problem, for simplicity, let us consider the example of QKD. The first QKD protocol is published in 1984 by Bennett and Brassard (BB84)~\cite {Bennett:BB84:1984}. Lots of progress has been made since then both theoretically and experimentally~\cite {Scarani:QKDrev:2008}. For the BB84 protocol, secure key bits can be transmitted only when single photon states are used. From the study of photon-number-splitting (PNS) attacks~\cite {BLMS:PNS:2000}, one can see that multi photon components are not secure for the BB84 protocol.  The key idea to take this imperfection into consideration is by performing privacy amplification on key bits from good (single photon) states and bad (multi photon) states separately~\cite {GLLP:2004}. Meanwhile, in order to accurately quantify the amounts of key bits from good and bad states, the decoy-state method has been proposed~\cite{Hwang:Decoy:2003,Lo:Decoy:2005,Wang:Decoy:2005} and experimentally demonstrated~\cite {Zhao:DecoyExp:2006, Rosenberg:ExpDecoy:2007,PDC144_07, Peng:ExpDecoy:2007, YSS_Decoy_07}.

For all the existing security analysis for coherent-state QKD protocols, including a recent QKD protocol~\cite {newQKD}, continuous phase randomization, Eq.~\eqref{Discrete:ContinuousPhase}, is assumed.
It has been shown that when the phase is not randomized, the performance will be substantially reduced with a strict security proof~\cite {LoPreskill:NonRan:2007}. In fact, there are experimental quantum hacking demonstrations showing that a QKD system may be attacked when the phase is not randomized lately~\cite {Sun:Partially:2012,tang2013source}.

There are two means to randomize the phase in practice, passive and active. In a passive phase randomization process, the laser is turned on and off to generate pulses. One might be tempted to make a naive argument that by switching a laser on and off, the phase is fully randomized. Note that it is experimentally challenging to verify rigorously that a {\it continuous} phase is indeed {\it fully} random. Moreover, experiments in random number generation have shown that there are indeed residue correlations between the phases of adjacent pulses~\cite {Xu:QRNG:2012}, especially in the case of high-speed applications~\cite {Abellan:14}, which directly rejects the claim since fully randomized phases have no correlations. Thus we avoid this approach here.

In the active phase randomization process, a phase modulator is used to randomly modulate the phases. In this case, the modulator can only perform discrete phase randomization, unless it uses an infinite amount of random numbers. In a recent experiment~\cite {Zhiyuan2013}  with coherent states, each global phase is chosen from one of the over 1000 possible values. First, such a large number of phases demands high precision control, which is a challenge for practical implementations. Second, even with 1000 phases, the phase is still discrete and the key rate may be deviated from the continuous-randomization case. So far,  \emph{no} rigorous bound on the key rate was derived in this experiment~\cite {Zhiyuan2013}  or any other QKD experimental papers. Since the work of Lo and Preskill was published~\cite {LoPreskill:NonRan:2007}, it has been a long-standing question to analyze the security of a practical QKD system with discrete-phase randomization and rectify the highly unsatisfactory situation that there is no proof of security in existing experiments.

In this work, we solve this long-standing open question by providing a rigorous security proof of QKD systems using discrete-phase-randomized coherent states. Here, we consider unconditional security, following the standard security proof \cite{GLLP:2004}. That is to say, security against the most general type of attacks allowed by quantum mechanics on the quantum channel by an eavesdropper.
We show that as the number of discrete phases increases, coherence between the Fock states in Eq.~\eqref{Discrete:CoherentState} decreases exponentially fast. As an application, we provide tight security bounds for both non-decoy and decoy state QKD protocols with discrete phase randomization. Our result applies to various encoding schemes of QKD including time-bin, phase encoding and polarization encoding.

In simulation, we compare the performance of our security bounds with the one provided by continuous phase randomization, which shows that our security bounds are tight when the number of phases goes to infinity. From a practical point of view, for small number of phases (say, only $N=10$ phases), with a typical set of experimental parameters, we observe that secret keys can be securely distributed over a fiber length of up to $138$ km, close to $140$ km in the continuous phase randomized case. Thus Alice needs only less than 4 bits ($2^4>10$) of random numbers per pulse for phase randomization. In contrast, all previous security proofs essentially assume an infinite number of bits of random numbers per pulse. Therefore, we are making an improvement here. Moreover, our scheme is simple to implement. For instance, an implementation of active phase randomization with 1000 discrete phases has been reported in literature~\cite {Zhiyuan2013}. Due to the massive reduction in the number of phases in our scheme (10 phases), one can expect a simpler implementation with a much higher repetition rate. Besides QKD protocols, our analysis of discrete phase randomization is also readily applicable to linear optics quantum computation~\cite {KLM_01} and other quantum cryptographic primitives~\cite {PhysRevA.84.052305}, because phase-randomized coherent sources serve also as major parts in them.

\section{Results}
The roadmap of this section is as follows. In Section~\ref{Sec:MixCoherent}, we use the Schmidt decomposition to construct states that are close to Fock states from discrete-phase-randomized coherent states. In Section~\ref{Sec:Scheme}, we present the phase encoding scheme when discrete-phase-randomized coherent states are used, and investigate how close the approximated Fock states are. In Section~\ref{sec:phaseerr}, we give a security proof and derive a key rate formula for a QKD system with discrete-phase randomization. In Section~\ref{sec:para}, we present the simulation results for two cases: with and without the decoy-state method.

\subsection{Coherent state mixture}
\label{Sec:MixCoherent}
Here, we consider a coherent state source whose phase is randomly picked from $N$ different values, i.e., each with probability $1/N$. For the sake of simplicity, we assume these $N$ values are evenly distributed in $[0,2\pi)$,
\begin{equation} \label{Discrete:Nphases}
\{\theta_k=\frac{2\pi k}{N} | k=0,1,\dots,N-1\}.
\end{equation}
In the case of continuous phase randomization (when $N\rightarrow\infty$), Eq.~\eqref{Discrete:ContinuousPhase} essentially shows that one can decompose the phase-randomized coherent mixed state into a statistical mixture of Fock states, $\ket{n}\bra{n}$. In the application of QKD, as well as quantum computing~\cite {KLM_01}, the single photon state, $\ket{1}\bra{1}$, is the most
important component.

In the case of a finite $N$, one can decompose the mixed state to a set of pure states and hopefully, one of which is close to single photon state. First let us consider the case $N=2$. We start with the initial state
\begin{equation} \label{Discrete:2mixed}
\ket{\Psi_2} = \ket{0}_A\ket{\sqrt{2}\alpha}_B+\ket{1}_A\ket{-\sqrt{2}\alpha}_B,
\end{equation}
where the phase of coherent state $\ket{\sqrt{2}\alpha}_B$ is controlled by a quantum coin $A$. The factor
$\sqrt{2}$ is included in the state for system B to simplify later
discussions. The normalization factor is ignored throughout the paper unless it matters. By performing a Schmidt decomposition
\begin{equation} \label{Discrete:2SchmidtDec}
\ket{\Psi_2} = (\ket{0}_A+\ket{1}_A)\ket{\lambda_0}_B+(\ket{0}_A-\ket{1}_A)\ket{\lambda_1}_B,
\end{equation}
where the two pure states are given by
\begin{eqnarray} \label{Discrete:OddEvenrho}
\ket{\lambda_0} &=& \ket{\sqrt{2}\alpha}+\ket{-\sqrt{2}\alpha}, \\
\ket{\lambda_1} &=&\ket{\sqrt{2}\alpha}-\ket{-\sqrt{2}\alpha}. \nonumber
\end{eqnarray}
By substituting the definition of coherent state, Eq.~\eqref{Discrete:CoherentState}, it is not hard to see that $\ket{\lambda_0}$ ($\ket{\lambda_1}$) is a superposition of even (odd) photon number Fock states. By this decomposition, the Hilbert space is divided into the even and odd number Fock state spaces, $H_{even}\oplus H_{odd}$. Since $\ket{\lambda_1}$ only contains odd photon number Fock states, we expect it is close to a single photon state, which can be confirmed from the calculation of fidelity later.

In the case of general $N\ge1$, the decomposition is similar but a bit more complex,
\begin{eqnarray} \label{Discrete:Nrho}
\ket{\Psi_N} &=& \sum\limits_{k=0}^{N-1}\ket{k}_A\ket{\sqrt{2}\alpha e^{2k\pi i/N}}_B \\
&=& \sum\limits_{j=0}^{N-1} \ket{j}_A \ket{\lambda_j}_B \nonumber
\end{eqnarray}
where $\ket{j}_A$ can be understood as a quantum coin with $N$ random outputs and the $N$ pure states are given by
\begin{equation} \label{Discrete:modNrho}
\ket{\lambda_j} = \sum_{k=0}^{N-1} e^{-2kj\pi i/N}\ket{e^{2k\pi i/N}\sqrt{2}\alpha}. \\
\end{equation}
By substituting Eq.~\eqref{Discrete:CoherentState}, we have the following observations for $\ket{\lambda_j}$. It is a superposition of Fock states whose photon numbers modulo $N$ are the same $j$,
\begin{equation} \label{Discrete:rhojFock}
\ket{\lambda_j}=\sum_{l=0}^{\infty} \frac{(\sqrt{2}\alpha)^{lN+j}}{\sqrt{(lN+j)!}}\ket{lN+j}. \\
\end{equation}
Then, it is not hard to see that $\ket{\lambda_j}$ becomes close to a Fock state when $N$ is large since $\sqrt{(lN+j)!}$ increases fast. When $N\rightarrow\infty$, it becomes a Fock state $\ket{\lambda_j}=\ket{j}$. Later in the simulation, one can see that when $N=10$, the mixed coherent state becomes close to Fock state mixture in terms of performance of QKD. Similar to the case of $N=2$, the Hilbert space is divided into $H_{0~\bmod~N}\oplus H_{1~\bmod~N}\oplus\dots\oplus H_{(N-1)~\bmod~N}$.

Next, we can figure out the probability if Alice performs a projection measurement on the photon state in the basis of $\ket{\lambda_j}$, which is simply the norm of Eq.~\eqref{Discrete:rhojFock},
\begin{eqnarray} \label{Discrete:Probj}
P_j &=& \frac{\braket{\lambda_j}{\lambda_j}}{\sum_{j=0}^{N-1} \braket{\lambda_j}{\lambda_j}} \nonumber \\
&=& \sum_{l=0}^{\infty}\frac{\mu^{lN+j}e^{-\mu}}{(lN+j)!},
\end{eqnarray}
where $\mu=2|\alpha|^2$. When $N\rightarrow\infty$, it becomes a photon number channel and follows a Poisson distribution $\mu^j e^{-\mu}/j!$.

\subsection{Coherent state scheme}
\label{Sec:Scheme}
\begin{figure}[htb]
\centering \includegraphics[width=12cm]{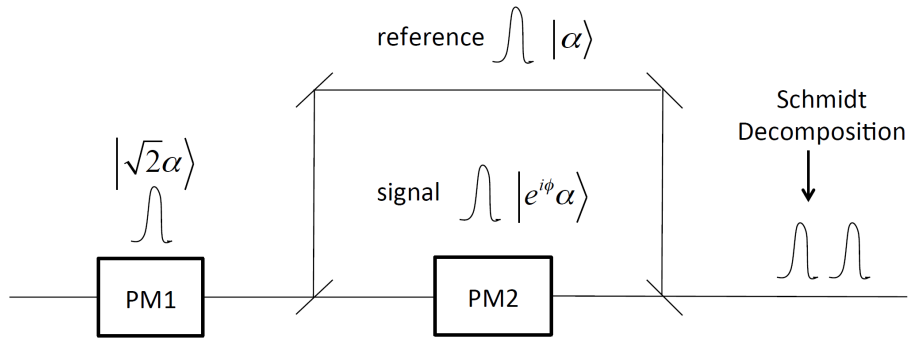}
\caption{Schematic diagram for the phase-encoding QKD scheme with coherent states. The first phase modulator, $PM1$, is used for phase randomization according to Eq.~\eqref{Discrete:Nphases}, and the second one, $PM2$, is used for QKD encoding $\phi\in\{0,\pi/2,\pi,3\pi/2\}$.} \label{Fig:ExpScheme}
\end{figure}

A typical scheme using a coherent state source, e.g., a phase encoding QKD scheme, is shown in Fig.~\ref{Fig:ExpScheme}, which is essentially an interferometer. In the state preparation stage, Alice prepares a weak coherent state $\ket{\sqrt{2}\alpha}$, whose phase is modulated randomly by the first phase modulator $PM1$. The state is separated into two pulses, $\ket{\alpha}_r$ and $\ket{\alpha}_s$, by a beam splitter. And then Alice encodes the bit and basis information (say, according to the BB84 protocol) in the relative phase via the second phase modulator $PM2$.

Here, for simplicity, we consider the case that the reference pulse has the same intensity with the signal. Our results can be extended to the strong reference case~\cite{Koashi_StrRef_04, Tamaki:Strong:08} and the asymmetric case~\cite{Norbert:Strong:2012}, as well as other encoding schemes such as polarization encoding and time-bin encoding~\cite{marcikic2002time}.

In the scheme with discrete $N$-phase randomization, the photon source is decomposed into states $\ket{\lambda_j}$ as shown in Eq.~\eqref{Discrete:modNrho}. After going through the phase encoding scheme as shown in Fig.~\ref{Fig:ExpScheme}, the four BB84 states encoded in $\ket{\lambda_j}$ can be written as
\begin{eqnarray} \label{Discrete:BB84logic4}
\ket{0_x^L} &=& \sum_{k=0}^{N-1} e^{-2kj\pi i/N}\ket{e^{2k\pi i/N}\alpha} \ket{e^{2k\pi i/N}\alpha} \nonumber \\
\ket{1_x^L} &= &\sum_{k=0}^{N-1} e^{-2kj\pi i/N}\ket{e^{2k\pi i/N}\alpha} \ket{-e^{2k\pi i/N}\alpha} \nonumber \\
\ket{0_y^L} &= &\sum_{k=0}^{N-1} e^{-2kj\pi i/N}\ket{e^{2k\pi i/N}\alpha} \ket{ie^{2k\pi i/N}\alpha} \\
\ket{1_y^L} &= &\sum_{k=0}^{N-1} e^{-2kj\pi i/N}\ket{e^{2k\pi i/N}\alpha} \ket{-ie^{2k\pi i/N}\alpha},\nonumber
\end{eqnarray}
where we omit the subscript $j$ on the left side, but it should be understood that the four states do depend on $j$.

The key point to guarantee the security of the BB84 protocol is that Eve cannot distinguish the state in two conjugate bases, $X$ and $Y$. The two density matrices in the two bases can be written as
\begin{eqnarray} \label{Discrete:rhoxy}
\rho_x &= &\ket{0_x^L}\bra{0_x^L}+\ket{1_x^L}\bra{1_x^L} \nonumber \\
\rho_y &=& \ket{0_y^L}\bra{0_y^L}+\ket{1_y^L}\bra{1_y^L}.
\end{eqnarray}
Note that each logical state should be regarded as a pure normalized state. In the ideal case, where a basis-independent source, such as a single photon source, is used, the density matrices in the two bases should be the same,
\begin{equation} \label{Discrete:BasisInd}
\rho_x = \rho_y.
\end{equation}
In the security analysis, one of the key parameters is the basis dependence of the source, which is the fidelity between the two states in the $X$ and $Y$ bases,
\begin{eqnarray}
&F_j(\rho_x,\rho_y) =\mathrm{tr} \sqrt{\sqrt{\rho_y} \rho_x \sqrt{\rho_y}} \label{Discrete:FidelityXY}\\
&\ge \left| \frac{\sum_{l=0}^{\infty} \frac{\mu^{lN+j}}{(lN+j)!} 2^{-\frac{lN+j}{2}} \left(\cos\frac{lN+j}{4}\pi+\sin\frac{lN+j}{4}\pi\right) }{\sum_{l=0}^{\infty}\frac{\mu^{lN+j}}{(lN+j)!}} \right| \nonumber
\end{eqnarray}
where $\mu=2|\alpha|^2$ and the detailed fidelity evaluation is shown in the Appendices.

Denote $F^{(t)}_j$ as the $t$-th order approximation of the fidelity, by taking $l=0,\dots, t$ in the summation.  The zeroth order is
\begin{equation} \label{Discrete:F0thj}
F^{(0)}_j \ge \left| 2^{-j/2} \left(\cos\frac{j}{4}\pi+\sin\frac{j}{4}\pi\right) \right| +O\left(\frac{\mu^{N}j!}{(N+j)!}\right).
\end{equation}
One can see that $F^{(0)}_0=F^{(0)}_1=1$ and $F^{(0)}_2=1/2$, $F^{(0)}_3=0$, $F^{(0)}_4=1/4$, $\dots$. Since when $F<1/\sqrt2$ would not render any positive key rate~\cite {LoPreskill:NonRan:2007}, it is confirmed that multi photon states are not secure for QKD due to their large basis dependence in the BB84 protocol.

\begin{figure}[hbt]
\centering \resizebox{12cm}{!}{\includegraphics{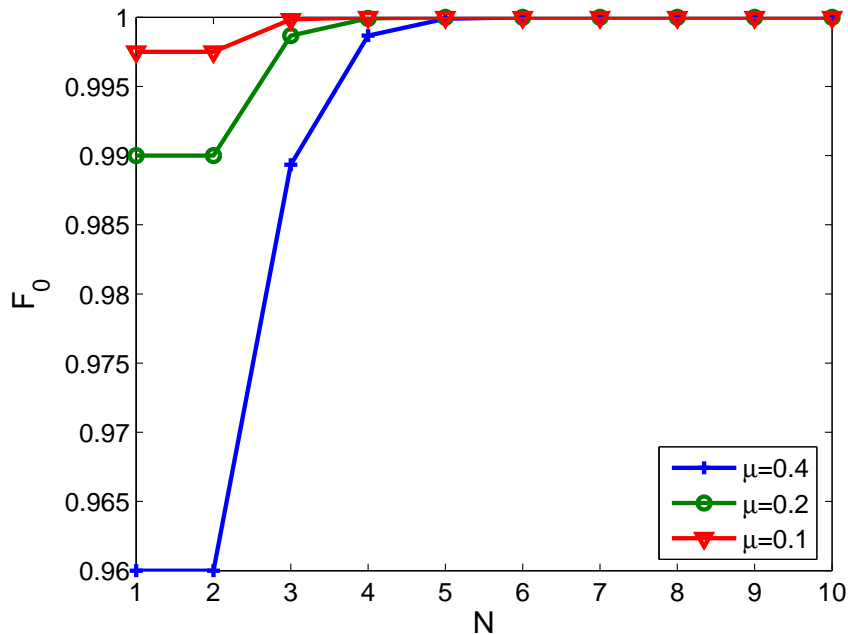}}
\caption{(Color Online) Plots of fidelity for different values of mean photon number $\mu$. The fidelity here refers to $F_0$, which is always smaller than $F_1$ by the numerical evaluation.} \label{Fig:fid}
\end{figure}

Take the first order for $\ket{\lambda_0}$ and $\ket{\lambda_1}$,
\begin{eqnarray} \label{Discrete:F1st01}
F^{(1)}_0 &\ge &1 -\left(1-2^{-\frac{N-1}{2}}\cos\frac{N-1}{4}\pi\right)\frac{\mu^N}{N!} +O\left(\frac{\mu^{2N}}{(N!)^2}\right) \\
F^{(1)}_1 &\ge& 1 -\left(1-2^{-\frac{N}{2}} \cos\frac{N}{4}\pi \right)\frac{\mu^N}{(N+1)!} +O\left(\frac{\mu^{2N}}{[(N+1)!]^2}\right). \nonumber
\end{eqnarray}
The fidelity approaches to 1 rapidly as $N$ becomes large, especially when $\mu$ is small, as shown in Fig.~\ref{Fig:fid}.
This shows with enough discrete phases, one can approximate the vacuum state and the single photon state infinitely well, which is useful in applications such as QKD.

\subsection{Key rate}
\label{sec:phaseerr}
In the perfect phase randomized case, the key rate formula is given by the standard GLLP security analysis for QKD with practical devices~\cite{GLLP:2004,Lo:Decoy:2005},
\begin{eqnarray} \label{Discrete:GLLPkey}
R &\ge& -I_{ec}+Q_{1}[1-H(e_1^p)], \nonumber \\
I_{ec} &=& fQ_\mu H(E_\mu), \\
Q_1 &=& Y_1\mu e^{-\mu}.  \nonumber
\end{eqnarray}
Here, $I_{ec}$ is the cost of error correction; $Q_{\mu}$ and $E_{\mu}$ are the overall gain and quantum bit error rate (QBER), respectively, which can be directly measured in QKD experiments; $Q_1$, $Y_1$, and $e_1^p$ are the gain, yield, and phase error rate of the single-photon component, respectively; $\mu e^{-\mu}$ is the (Poisson) probability that Alice sends single photon states; $\mu$ denotes the expected photon number of the signal state; $f$ denotes the error correction efficiency; and $H(p)=-p\log_{2}(p)-(1-p)\log_{2}(1-p)$ is the binary Shannon entropy function, where $p$ is a binary probability. We assume that Alice and Bob run the efficient BB84~\cite {Lo:EffBB84:2005} and take the basis sift factor to be 1.

The QKD key rate formula, Eq.~\eqref{Discrete:GLLPkey}, is derived using ideas of entanglement distillation \cite{LoChauQKD_99} and complementarity \cite{Koashi_Uncer_06}. It satisfies the composable security definition \cite{BenOr:Security:05, Renner:Security:05}. In QKD protocols, Alice and Bob need to perform error correction to eliminate the errors and share an identical key. In this error correction procedure, a fraction of $I_{ec}$ sacrificed from the raw key. Then, they need to eliminate the eavesdropper's information on the error-corrected key via privacy amplification. The perfect phase randomization allows us to consider the signal as a mixture of Fock states and estimate the contributions of the components of different photon numbers separately~\cite{Lo:Decoy:2005}. Since multi-photon components are not secure in the BB84 protocol~\cite{BLMS:PNS:2000}, only the single-photon component $Q_1$ will appear in the key formula \cite{GLLP:2004}. The amount of the eavesdropper's information in the single-photon component is related to the phase error rate, $e_1^p$.

The core of a practical security analysis is to figure out the privacy amplification term $Q_{1}[1-H(e_1^p)]$ in Eq.~\eqref{Discrete:GLLPkey}. For a single-photon state, it is a basis-independent source, thus its phase error rate is equal to its bit error rate \cite{ShorPreskill_00}. Now, the key point of the analysis is to estimate the yield and bit error rate of the single-photon component, $Y_1$ and $e_1^b$. This estimation can be done with different means, such as the decoy-state method \cite{Hwang:Decoy:2003,Lo:Decoy:2005,Wang:Decoy:2005}.


In the case of discrete phase randomization, the photon source is not decomposed into Fock states. Instead, we decompose the channel into $\ket{\lambda_j}$ according to Eq.~\eqref{Discrete:Nrho}. The single photon state will be replaced by $\ket{\lambda_1}$ and the Poisson distribution will be replaced by Eq.~\eqref{Discrete:Probj}. Then, the approximated single-photon state $\ket{\lambda_1}$ is no longer a basis-independent source. The basis dependence of the source is evaluated in Section~\ref{Sec:Scheme}, which causes deviation between the bit and phase error rates.

Now we can slightly modify Eq.~\eqref{Discrete:GLLPkey} to fit our case
\begin{equation} \label{Discrete:GLLPmodif}
R\ge -I_{ec}+\sum_jP_j Y_j[1-H(e_j^p)],
\end{equation}
where $P_j$ is given in Eq.~\eqref{Discrete:Probj}. The yield $Y_j$ and bit error rate $e_j^b$  of $\ket{\lambda_j}$, can be estimated by the decoy state method. Here, without any confusion, we use the same notation as the Fock state case for simplicity. Given the basis dependence $\Delta_j$, one can bound the phase error rate $e_j^p$ from $e_j^b$ similar to the work of Lo and Preskill \cite{LoPreskill:NonRan:2007},
\begin{equation} \label{Discrete:bitphase}
e_j^p \le e_j^b+4\Delta_j(1-\Delta_j)(1-2e_j^b)+4(1-2\Delta_j)\sqrt{\Delta_j(1-\Delta_j) e_j^b(1-e_j^b)}.
\end{equation}
The basis dependence is defined as
\begin{equation} \label{Discrete:bias}
\Delta_j = \frac{1-F_j}{2Y_j}.
\end{equation}
where the fidelities $F_j$ are given in Eq.~\eqref{Discrete:FidelityXY}. The key difference between our result and the original GLLP analysis is that the bit and phase error rates are not the same in the ``single"-photon component.

From the evaluation of the basis dependence, it is not hard to show that only $j=0$ and $j=1$ would contribute positively to the final key rate. Thus, the key rate evaluation becomes the following minimization problem.
\begin{equation} \label{Discrete:minKey}
\min_{0\le Y_j,e_j^b\le1} \{P_0 Y_0[1-H(e_0^p)] +P_1 Y_1[1-H(e_1^p)]\}.
\end{equation}
There are other constraints based on the gain and QBER obtained from the experiments. Note that with other security proof techniques the key rate given in Eq.~\eqref{Discrete:GLLPmodif} can be improved. For example, the vacuum component $Q_0$ is showed~\cite {Lo:Vacuum:2005} to have no phase errors when the photon number channel model is applied.

\subsection{Parameter estimation}
\label{sec:para}
Now, we need to estimate the key parameters, $Y_j$ and $e_j^b$. First, let us consider the no-decoy state case, where we assume all the losses and errors come from $\ket{\lambda_0}$ and $\ket{\lambda_1}$, in the worst case scenario,
\begin{eqnarray} \label{Discrete:noDecoyYe}
P_0 Y_0 +P_1 Y_1 \ge Q_\mu-\sum_{j=2}^{N-1} P_j, \nonumber \\
e_0P_0 Y_0 +e_1P_1 Y_1 \le E_\mu Q_\mu.
\end{eqnarray}
Since the right side of Eq.~\eqref{Discrete:noDecoyYe} can be obtained from the experiment directly, one can easily solve the minimization problem presented in Eq.~\eqref{Discrete:minKey} to get the key rate.

\begin{figure}[hbt]
\centering \resizebox{12cm}{!}{\includegraphics{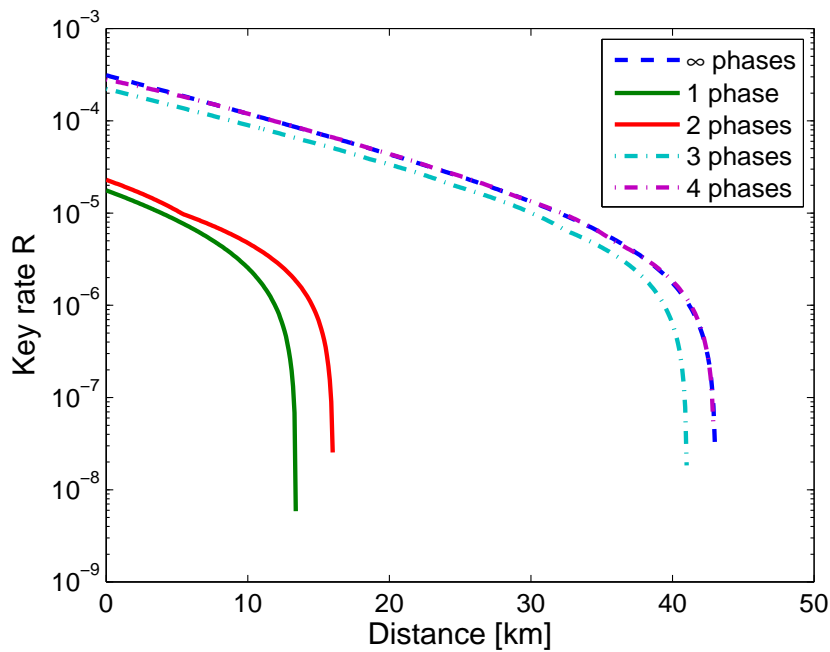}}
\caption{(Color Online) Plots of key rates of the no-decoy state scheme for different numbers of random phases $N$. When $N=1$, it refers to no phase randomization case. When $N\rightarrow\infty$, it approaches to continuous phase randomization case.} \label{Fig:nodecoy}
\end{figure}

We simulate a typical QKD system~\cite {GYS:122km:2004} and compare various cases of $N$.
The result is shown in Fig.~\ref{Fig:nodecoy}, from which we can see that with only 4 random phases, the performance of discrete phase randomization is close to the one of continuous phase randomization. We can also observe the key rate of one phase and two phases are similar, but there is a gap when the phase number becomes three. This can be explained as follows. We note that $F_0^{(2)}$ of $N=1$ coincides with Eq. (22) in the work of Lo and Preskill~\cite {LoPreskill:NonRan:2007}, thus our fidelity formula also extends to the $N=1$ case. Also we notice that the first order term of $N=1$ vanishes, making the key rate performance of $N=1$ and $N=2$ to be similar, both of order $1-O(\mu^2)$. For $N\ge 3$, the performance is improved  to $1-O(\mu^N)$.
The details of this simulation and all following simulations are shown in the Appendices.

For the case of the decoy-state method, the analysis is trickier. In the perfect phase randomized case, the decoy state method immensely improves the key rate by offering accurate parameter estimation for Eq.~\eqref{Discrete:minKey}. In the security proof of the decoy-state method, the photon number channel model guarantees the following equalities,
\begin{eqnarray} \label{Discrete:DecoyAssum}
Y_n(signal) &= Y_n(decoy), \nonumber \\
e_n(signal) &= e_n(decoy)
\end{eqnarray}
since all the Fock states $\ket{n}\bra{n}$ are the same in the signal and decoy states.
Thus adding decoy states imposes more equations constraints on the parameters,
\begin{eqnarray} \label{Continuous:QESD}
Q_\mu &=& \sum_{j=0}^{+\infty} \frac{\mu^j e^{-\mu}}{j!} Y_j, \nonumber \\
E_\mu Q_\mu &=& \sum_{j=0}^{+\infty}  \frac{\mu^j e^{-\mu}}{j!}e_j  Y_j, \nonumber \\
Q_\nu &=& \sum_{j=0}^{+\infty}   \frac{\nu^j e^{-\nu}}{j!} Y_j, \\
E_\nu Q_\nu &=& \sum_{j=0}^{+\infty} \frac{\nu^j e^{-\nu}}{j!} e_j  Y_j, \nonumber
\end{eqnarray}
without inducing more variables other than the original unknown variables.
Here, $\mu^j e^{-\mu}/j!$ is the (Poisson) probability that Alice sends $j$ photons states.
This consequently gives tighter bounds on $Y_j$ and $e_j$.

This is not so straightforward in $N$ discrete phase randomization case, because
\begin{equation} \label{Discrete:AssumFail}
\ket{\lambda_j^\mu} \neq \ket{\lambda_j^\nu}
\end{equation}
as defined in Eq.~\eqref{Discrete:rhojFock}, where $\mu$ and $\nu$ are the intensities of signal and decoy states. Thus, we do not have the simple relations as the continuous phase randomization case, Eq.~\eqref{Discrete:DecoyAssum}. Fortunately, we have shown that $\ket{\lambda_j}$ is close to the Fock state $\ket{j}$. We expect the inequality shown in Eq.~\eqref{Discrete:AssumFail} to be an approximate equality.

Following the quantum coin argument used in the GLLP security analysis~\cite {GLLP:2004}, the yield and error rate difference between the signal and decoy states are given by
\begin{eqnarray} \label{Discrete:YSDdiff}
|Y_j^\mu-Y_j^\nu| &\le& \sqrt{1-F_{\mu\nu}^2}, \\
|e_j^\mu Y_j^\mu-e_j^\nu Y_j^\nu| &\le& \sqrt{1-F_{\mu\nu}^2}, \nonumber
\end{eqnarray}
where
\begin{equation} \label{Discrete:Fmunu}
F_{\mu\nu} =  1- O\left(\frac{\mu^{N}}{N!}\right).
\end{equation}

Now the extra constraints added to the minimization problem of Eq.~\eqref{Discrete:minKey} for the decoy-state method are, along with Eq.~\eqref{Discrete:YSDdiff},
\begin{eqnarray} \label{Discrete:QESD}
Q_\mu &=& \sum_{j=0}^{N-1} P_j^\mu Y_j^\mu, \nonumber \\
E_\mu Q_\mu &=& \sum_{j=0}^{N-1} e_j^\mu P_j^\mu Y_j^\mu, \nonumber \\
Q_\nu &=& \sum_{j=0}^{N-1} P_j^\nu Y_j^\nu, \\
E_\nu Q_\nu &=& \sum_{j=0}^{N-1} e_j^\nu P_j^\nu Y_j^\nu, \nonumber
\end{eqnarray}
where $P_j^\mu$ are given in Eq.~\eqref{Discrete:Probj}. If more decoy states are used, more linear equations will be added to Eq.~\eqref{Discrete:QESD}.

\begin{figure}[hbt]
\centering \resizebox{12cm}{!}{\includegraphics{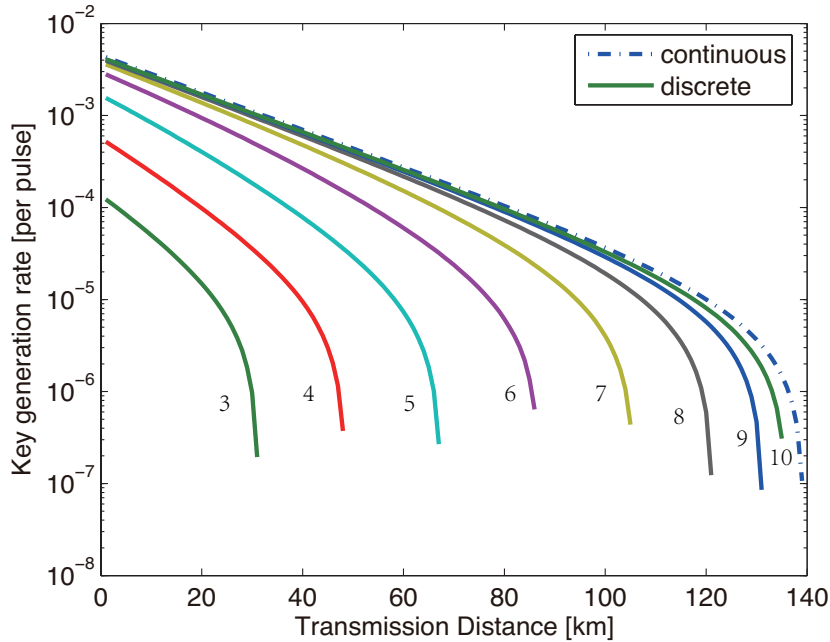}}
\caption{(Color Online) Plots of key rates of the decoy-state scheme for different numbers of random phases $N$.  Dashed line refers to the continuous phase randomization case. Solid lines from left to right refer to increasing $N$ from 3 to 10.} \label{Fig:decoy}
\end{figure}

We simulate a QKD system~\cite {GYS:122km:2004} with vacuum+weak decoy state~\cite {MXF:Practical:2005} and compare various cases of phase number $N$.  The decoy and signal intensities are numerically optimized to maximize the key rate. The result is shown in Fig.~\ref{Fig:decoy}, from which we can see that with only 10 random phases, the performance of discrete phase randomization is close to the one of continuous phase randomization.

Notice that the QBER with discrete phases highly depends on the experimental parameters. According to our simulation model, each QBER in Fig.~\ref{Fig:nodecoy} and Fig.~\ref{Fig:decoy} is a function of the signal intensity $\mu$, which is numerically optimized to maximize the key rate. This may not lead to the highest tolerable QBER for a given transmission distance. The reason is that by tuning the signal intensity $\mu$ smaller, one can increase the tolerable QBER at the expense of lowering the key rate. Eventually, the hard bound on the allowable QBER is 11\% just like the single photon BB84 protocol despite the number of discrete phases used, as we are following the Shor-Preskill security proof \cite{ShorPreskill_00} and an infinitely small $\mu$ effectively turns a coherent state source into a single photon source.

\section{Discussion}

In summary, we just need 10 random phases for the discrete phase randomization, the fidelity of which is close to
the continuous  case. We demonstrate the effect of discrete phase randomization by taking the QKD protocol as an example and show that it gives a big improvement on the performance.
Without phase randomization, the key rate decays rapidly as a function of the transmittance of
the channel and drops to zero after less than 15km of optical fibers as shown in
Fig.~\ref{Fig:nodecoy}. In contrast, with discrete phase randomization,
the key rate scales linearly as a function of
the transmittance and QKD remains feasible over 138km of
fibers as shown in Fig.~\ref{Fig:decoy}. Since only four bits of random numbers per pulse, which already give $2^4=16>10$ possible phases, are required for phase randomization, our scheme is highly practical. Note that a much harder discrete phase randomization experiment with 1000 phases~\cite {Zhiyuan2013}  has already been demonstrated.
Moreover, our method may not only  apply to the same signal and reference pulse amplitudes case, but also to the asymmetric amplitude case and the strong reference pulse case. We remark that our discrete phase randomization idea applies  to other quantum information processing protocols including blind quantum computing and quantum coin tossing.

There are a few interesting prospective projects. First, due to the finite length of the key, statistical fluctuation needs to be taken into consideration which can be dealt with by finite key analysis as in a recent work~\cite {PhysRevA.89.022307}. Second, $N$-discrete-phase-randomization process might not be perfect in an actual system, i.e., there can be a small fluctuation in the phase modulation such that the actual phase applied will be $\{\delta_k + \frac{2 \pi k}{ N}\}_{k=0,1,\ldots,n-1}$ where $\delta_k$ is a small fluctuating value  that can be positive or negative. The imperfect phase modulation can be dealt with by modifying our fidelity calculations. More precisely, one can replace the coherent state $\ket{e^{2k\pi i/N}\alpha}$ by $\ket{e^{2k\pi i/N+\delta_k}\alpha}$ in Eq.~\eqref{Discrete:BB84logic4}  and calculate the fidelity in Eq.~\eqref{Discrete:FidelityXY}. We expect the result will be robust against small $\delta_k$. Besides the usual BB84 QKD protocols, our idea can also be extended to measurement-device-independent QKD~\cite{Lo:MDIQKD:2012} by treating both sources as discrete phase randomized.

\section*{Acknowledgements}
The authors acknowledge insightful discussions with C.-H.~F.~Fung, and X.~Yuan. This work was supported by the National Basic Research Program of China Grants No.~2011CBA00300 and No.~2011CBA00301, the 1000 Youth Fellowship program in China and NSERC.

\appendix

\section{Basis dependence} \label{Sec:FidelityXY}
This appendix and the following appendices are functioned as follows. In \ref{Sec:FidelityXY}, the basis dependence between the $X$ and $Y$ (fidelity of two density matrices) for $\ket{\lambda_j}$ is calculated when $N$ discrete randomized phases are used. In \ref{AppSec:DecoyEst}, we present the parameter estimation of the decoy-state method. In \ref{AppSec:Sim}, the pseudo codes for both nondecoy and decoy simulation are given.

In order to make the derivation easier to understand, we use index $n$ to represent the photon number, index $k$ to represent the discrete phase, index $j$ to represent the decomposed Fock state approximations.

We restate the four BB84 states phase encoded in the decomposed state $\ket{\lambda_j}$ as presented in Main Text,
\begin{eqnarray} \label{App:BB84logic4}
\ket{0_x^L} &=& \frac{\sum_{k=0}^{N-1} e^{-2kj\pi i/N}\ket{e^{2k\pi i/N}\alpha} \ket{e^{2k\pi i/N}\alpha}}{\sqrt{Ne^{-2|\alpha|^2}\sum_{k=0}^{N-1} e^{2kj\pi i/N}e^{2|\alpha|^2e^{-2k\pi i/N}}}} \nonumber \\
\ket{1_x^L} &=& \frac{\sum_{k=0}^{N-1} e^{-2kj\pi i/N}\ket{e^{2k\pi i/N}\alpha} \ket{-e^{2k\pi i/N}\alpha}}{\sqrt{Ne^{-2|\alpha|^2}\sum_{k=0}^{N-1} e^{2kj\pi i/N}e^{2|\alpha|^2e^{-2k\pi i/N}}}} \\
\ket{0_y^L} &=& \frac{\sum_{k=0}^{N-1} e^{-2kj\pi i/N}\ket{e^{2k\pi i/N}\alpha} \ket{ie^{2k\pi i/N}\alpha}}{\sqrt{Ne^{-2|\alpha|^2}\sum_{k=0}^{N-1} e^{2kj\pi i/N}e^{2|\alpha|^2e^{-2k\pi i/N}}}} \nonumber \\
\ket{1_y^L} &=& \frac{\sum_{k=0}^{N-1} e^{-2kj\pi i/N}\ket{e^{2k\pi i/N}\alpha} \ket{-ie^{2k\pi i/N}\alpha}}{\sqrt{Ne^{-2|\alpha|^2}\sum_{k=0}^{N-1} e^{2kj\pi i/N}e^{2|\alpha|^2e^{-2k\pi i/N}}}}, \nonumber
\end{eqnarray}
where the denominators are the normalization factors.

To evaluate the fidelity between the two states in the two bases, we calculate the related inner products of these four states,
\begin{eqnarray} \label{App:Inner}
\braket{0_x^L}{0_y^L}&=&\braket{1_x^L}{1_y^L}
= \frac{Ne^{-2|\alpha|^2}\sum_{k=0}^{N-1} e^{2kj\pi i/N}e^{|\alpha|^2(1+i)e^{-2k\pi i/N}}}{Ne^{-2|\alpha|^2}\sum_{k=0}^{N-1} e^{2kj\pi i/N}e^{2|\alpha|^2e^{-2k\pi i/N}}}, \nonumber \\
\braket{0_x^L}{1_y^L}&=& \braket{1_x^L}{0_y^L}
= \frac{Ne^{-2|\alpha|^2}\sum_{k=0}^{N-1} e^{2kj\pi i/N}e^{|\alpha|^2(1-i)e^{-2k\pi i/N}}}{Ne^{-2|\alpha|^2}\sum_{k=0}^{N-1} e^{2kj\pi i/N}e^{2|\alpha|^2e^{-2k\pi i/N}}},
\end{eqnarray}

The detailed calculations of inner products and norms are shown in Section \ref{AppSub:InnerNorm}. Now we substitute these values to evaluate fidelity,
\begin{eqnarray} \label{App:Fidelity}
F(\rho_x,\rho_y) &=& F\left( \ket{0_x^L}\bra{0_x^L}+\ket{1_x^L}\bra{1_x^L}, \ket{0_y^L}\bra{0_y^L}+\ket{1_y^L}\bra{1_y^L} \right) \nonumber \\
&\ge& F\left(\ket{-}\ket{0_x^L}+\ket{+}\ket{1_x^L}, \ket{+i}\ket{0_y^L}+\ket{-i}\ket{1_y^L} \right) \nonumber \\
&=& \frac12|(\bra{-}\bra{0_x^L}+\bra{+}\bra{1_x^L})(\ket{+i}\ket{0_y^L}+\ket{-i}\ket{1_y^L})|  \nonumber \\
&=& \frac{\sqrt2}{4}\left| \braket{0_x^L}{0_y^L} +i\braket{0_x^L}{1_y^L} +i\braket{1_x^L}{0_y^L} +\braket{1_x^L}{1_y^L} \right| \\
&=& \frac{\sqrt2}{2}\left| \frac{\sum_{k=0}^{N-1}e^{2kj\pi i/N} \left[e^{|\alpha|^2(1+i)e^{-2k\pi i/N}} +i e^{|\alpha|^2(1-i)e^{-2k\pi i/N}}\right] }{\sum_{k=0}^{N-1} e^{2kj\pi i/N}e^{2|\alpha|^2e^{-2k\pi i/N}}} \right|  \nonumber
\end{eqnarray}
where $\ket{\pm}$ and $\ket{\pm i}$ are the normalized eigenstates of the $X$ and $Y$ bases. The inequality comes from the fact that the fidelity of two mixed states is the maximal of the fidelity of all the purifications. Here, we use the intuition that two Bell states are the same,
\begin{equation} \label{App:2Bell}
\ket{+}\ket{-}+\ket{-}\ket{+}=\ket{+i}\ket{+i}+\ket{-i}\ket{-i}.
\end{equation}

Now, let us simplify Eq.~\eqref{App:Fidelity} and we expect it to be close to 1 when $N$ is large.
\begin{equation} \label{App:FidelityS1}
F(\rho_x,\rho_y) \ge \frac{\sqrt2}{2}\left| \frac{\sum x^{-j} \left[e^{|\alpha|^2(1+i)x} +i e^{|\alpha|^2(1-i)x}\right] }{\sum x^{-j} e^{2|\alpha|^2x}} \right| \\
\end{equation}
where the summation is taken over $x=1,e^{2\pi i/N},\dots,e^{2\pi i(N-1)/N}$, $N$ dots evenly distributed on the unit circle of the complex plane. Take the Taylor expansion of $\mu=2|\alpha|^2\ge0$ around 0,
\begin{eqnarray} \label{App:FTaylor}
F(\rho_x,\rho_y) &\ge& \left| \frac{\sum x^{-j} \sum_{n=0}^{\infty}\frac{(\mu x/\sqrt2)^n}{n!}(\cos\frac{n\pi}{4}+\sin\frac{n\pi}{4}) }{\sum x^{-j} \sum_{n=0}^{\infty}\frac{(\mu x)^n}{n!}} \right| \nonumber\\
&=& \left| \frac{\sum_{n=0}^{\infty} \frac{(\mu/\sqrt2)^n}{n!}(\cos\frac{n\pi}{4}+\sin\frac{n\pi}{4}) \sum x^{n-j} }{\sum_{n=0}^{\infty}\frac{\mu^n}{n!}\sum x^{n-j}} \right| \\
&= &\left| \frac{\sum_{l=0}^{\infty} \frac{\mu^{lN+j}}{(lN+j)!} 2^{-\frac{lN+j}{2}} \left(\cos\frac{lN+j}{4}\pi+\sin\frac{lN+j}{4}\pi\right) }{\sum_{l=0}^{\infty}\frac{\mu^{lN+j}}{(lN+j)!}} \right|. \nonumber
\end{eqnarray}
The the details of Taylor expansion and the calculation of $\sum x^{n-j}$ are shown in Section \ref{AppSub:TaylorSum}.

\subsection{Approximations: large $N$ or small $\mu$}
Here, we want to check the fidelity given in Eq.~\eqref{App:FTaylor} when $N$ is large or $\mu$ is small. Zeroth order, by taking $l=0$ in the summation,
\begin{eqnarray} \label{App:F0thj}
 F^{(0)}_j &=& \left| \frac{\frac{\mu^{j}}{j!} 2^{-\frac{j}{2}} \left(\cos\frac{j}{4}\pi+\sin\frac{j}{4}\pi\right) }{\frac{\mu^{j}}{j!}} \right| + O\left(\frac{\mu^{N}j!}{(N+j)!}\right) \nonumber \\
&\approx& \left| 2^{-j/2} \left(\cos\frac{j}{4}\pi+\sin\frac{j}{4}\pi\right) \right|
\end{eqnarray}
One can see that $F^{(0)}_0=F^{(0)}_1=1$ and $F^{(0)}_2=1/2$, $F^{(0)}_3=0$, $F^{(0)}_4=1/4$, $\dots$. It is confirmed that multi photon states are not secure for the BB84 QKD protocol.

First order approximation, by taking $l=0$ and $l=1$ in the summation,
\begin{eqnarray} \label{App:F1stj}
F^{(1)}_j &= &\left| \frac{\frac{\mu^{j}}{j!} 2^{-\frac{j}{2}} \left(\cos\frac{j}{4}\pi+\sin\frac{j}{4}\pi\right) +\frac{\mu^{N+j}}{(N+j)!}2^{-\frac{N+j}{2}} \left(\cos\frac{N+j}{4}\pi+\sin\frac{N+j}{4}\pi\right)}{\frac{\mu^{j}}{j!} +\frac{\mu^{N+j}}{(N+j)!}} \right| + O\left(\frac{\mu^{2N}j!}{(2N+j)!}\right)\nonumber \\
&=& \left| \frac{ 2^{-\frac{j}{2}} \left(\cos\frac{j}{4}\pi+\sin\frac{j}{4}\pi\right) +\frac{\mu^N j!}{(N+j)!}2^{-\frac{N+j}{2}} \left(\cos\frac{N+j}{4}\pi+\sin\frac{N+j}{4}\pi\right)}{1 +\frac{\mu^N j!}{(N+j)!}} \right| + O\left(\frac{\mu^{2N}j!}{(2N+j)!}\right) \nonumber\\
&=& \left| 2^{-\frac{j}{2}} \left(\cos\frac{j}{4}\pi+\sin\frac{j}{4}\pi\right) -\left[1-2^{-\frac{N+j}{2}} \left(\cos\frac{N+j}{4}\pi+\sin\frac{N+j}{4}\pi\right) \right]\frac{\mu^N j!}{(N+j)!} \right| \nonumber\\
&\;&\;\;+ O\left(\left[\frac{\mu^{N}j!}{(N+j)!}\right]^2\right)+ O\left(\frac{\mu^{2N}j!}{(2N+j)!}\right)
\end{eqnarray}
Since $N\ge j$, the second $O(\cdot)$ in the last equality can be neglected. The first order approximation will approach to the zeroth order exponentially fast, $O(\mu^N/N!)$. We are interested in the first two cases $j=0$ and $j=1$,
\begin{eqnarray} \label{App:F1st01}
F^{(1)}_0 &=& 1 -\left[1-2^{-\frac{N}{2}} \left(\cos\frac{N}{4}\pi+\sin\frac{N}{4}\pi\right)\right]\frac{\mu^N}{N!} +O\left(\frac{\mu^{2N}}{(N!)^2}\right)\nonumber \\
&=& 1 -\left(1-2^{-\frac{N-1}{2}}\cos\frac{N-1}{4}\pi\right)\frac{\mu^N}{N!} +O\left(\frac{\mu^{2N}}{(N!)^2}\right) \\
F^{(1)}_1 &=& 1 -\left[1-2^{-\frac{N+1}{2}} \left(\cos\frac{N+1}{4}\pi+\sin\frac{N+1}{4}\pi\right)\right]\frac{\mu^N}{(N+1)!} +O\left(\left[\frac{\mu^{N}}{(N+1)!}\right]^2\right)  \nonumber \\
&=& 1 -\left(1-2^{-\frac{N}{2}} \cos\frac{N}{4}\pi \right)\frac{\mu^N}{(N+1)!} +O\left(\frac{\mu^{2N}}{[(N+1)!]^2}\right)\nonumber
\end{eqnarray}
We notes that the second order approximation of $F_0$ when $N=1$ coincides with Eq. (22) in~\cite {LoPreskill:NonRan:2007}, thus our fidelity formula also extends to $N=1$ case. Also we notice the first order term when $N=1$ vanishes, making the key rate performance of $N=1$ and $N=2$ to be similar, both of order $1-O(\mu^2)$. For $N\ge 3$, the performance is improved to $1-O(\mu^N)$.

\subsection{Inner products and norms} \label{AppSub:InnerNorm}
Inner products,
\begin{eqnarray} \label{App:Inner00}
\braket{0_x^L}{0_y^L} &=&\left(\sum_{l=0}^{N-1} e^{2lj\pi i/N}\bra{e^{2l\pi i/N}\alpha} \bra{e^{2l\pi i/N}\alpha}\right) \left(\sum_{k=0}^{N-1} e^{-2kj\pi i/N}\ket{e^{2k\pi i/N}\alpha} \ket{ie^{2k\pi i/N}\alpha}\right)\nonumber \\
&=& \sum_{l=0}^{N-1}\sum_{k=0}^{N-1} e^{2(l-k)j\pi i/N}\braket{e^{2l\pi i/N}\alpha}{e^{2k\pi i/N}\alpha} \braket{e^{2l\pi i/N}\alpha}{ie^{2k\pi i/N}\alpha} \nonumber\\
&= &\sum_{l=0}^{N-1}\sum_{k=0}^{N-1} e^{2(l-k)j\pi i/N}e^{-|\alpha|^2\left[2-(1+i)e^{2(k-l)\pi i/N}\right]} \\
&=& Ne^{-2|\alpha|^2}\sum_{k=0}^{N-1} e^{2kj\pi i/N}e^{|\alpha|^2(1+i)e^{-2k\pi i/N}} \nonumber
\end{eqnarray}
where we use the fact that $e^{2kj\pi i/N}$ and $e^{-2k\pi i/N}$ each forms a ring in the complex plane, and
\begin{eqnarray} \label{App:Inner01}
\braket{0_x^L}{1_y^L} &=& \left(\sum_{l=0}^{N-1} e^{2lj\pi i/N}\bra{e^{2l\pi i/N}\alpha} \bra{e^{2l\pi i/N}\alpha}\right) \left(\sum_{k=0}^{N-1} e^{-2kj\pi i/N}\ket{e^{2k\pi i/N}\alpha} \ket{-ie^{2k\pi i/N}\alpha}\right) \nonumber\\
&=& \sum_{l=0}^{N-1}\sum_{k=0}^{N-1} e^{2(l-k)j\pi i/N}\braket{e^{2l\pi i/N}\alpha}{e^{2k\pi i/N}\alpha} \braket{e^{2l\pi i/N}\alpha}{-ie^{2k\pi i/N}\alpha} \nonumber\\
&=& \sum_{l=0}^{N-1}\sum_{k=0}^{N-1} e^{2(l-k)j\pi i/N}e^{-|\alpha|^2\left[2-(1-i)e^{2(k-l)\pi i/N}\right]} \\
&= &Ne^{-2|\alpha|^2}\sum_{k=0}^{N-1} e^{2kj\pi i/N}e^{|\alpha|^2(1-i)e^{-2k\pi i/N}} \nonumber
\end{eqnarray}

Norms,
\begin{eqnarray} \label{App:Norm00}
\braket{0_x^L}{0_x^L} &=& \left(\sum_{l=0}^{N-1} e^{2l\pi i/N}\bra{e^{2l\pi i/N}\alpha} \bra{e^{2l\pi i/N}\alpha}\right) \left(\sum_{k=0}^{N-1} e^{-2k\pi i/N}\ket{e^{2k\pi i/N}\alpha} \ket{e^{2k\pi i/N}\alpha}\right) \nonumber\\
&= &\sum_{l=0}^{N-1}\sum_{k=0}^{N-1} e^{2(l-k)\pi i/N} \braket{e^{2l\pi i/N}\alpha}{e^{2k\pi i/N}\alpha}^2 \nonumber \\
&=& \sum_{l=0}^{N-1}\sum_{k=0}^{N-1} e^{2(l-k)\pi i/N}e^{-2|\alpha|^2\left[1-e^{2(k-l)\pi i/N}\right]} \\
&= &Ne^{-2|\alpha|^2}\sum_{k=0}^{N-1} e^{2k\pi i/N}e^{2|\alpha|^2e^{-2k\pi i/N}}\nonumber
\end{eqnarray}

Here, we use the inner products between two coherent states,
\begin{eqnarray} \label{App:CohOverlap}
\braket{\alpha}{\beta} &=& \exp\left( -\frac12|\alpha|^2+\alpha^*\beta-\frac12|\beta|^2 \right) \nonumber \\
\braket{\alpha e^{i\phi}}{\alpha e^{i\theta}} &=& e^{-|\alpha|^2(1-\exp[i(\theta-\phi)])}.
\end{eqnarray}
It is not hard to see that by adding a same phase to $\phi$ and $\theta$, the result is the same.

\subsection{Taylor expansion and summation} \label{AppSub:TaylorSum}
Taylor expansion:
\begin{eqnarray} \label{App:FTaylorUp}
e^{|\alpha|^2(1+i)x} +i e^{|\alpha|^2(1-i)x} &=& 1+\frac{1+i}{2}\mu x +\frac{(\frac{1+i}{2}\mu x)^2}{2!} +\frac{(\frac{1+i}{2}\mu x)^3}{3!} +\dots\nonumber \\
&\;&+i +i\frac{1-i}{2}\mu x +i\frac{(\frac{1-i}{2}\mu x)^2}{2!} +i\frac{(\frac{1-i}{2}\mu x)^3}{3!} +\dots \nonumber
\\
&= &(1+i)\sum_{n=0}^{\infty}(\frac{\mu x}{\sqrt2})^n\frac1{n!}(\cos\frac{n\pi}{4}+\sin\frac{n\pi}{4})
\end{eqnarray}

Summation:
\begin{eqnarray} \label{App:Fsumn}
\sum_x x^{n-j} &=& \sum_{k=0}^{N-1} e^{-2k(n-j)\pi i/N},
\end{eqnarray}
which equals $N$ if $n-j \textrm{ mod } N=0$ and equals $0$ if $n-j \textrm{ mod } N\neq 0$.
The summation is taken over $x=1,e^{2\pi i/N},\dots,e^{2\pi i(N-1)/N}$, $N$ dots evenly distributed on the unit circle of the complex plane.

\section{Parameter deviation in the decoy-state method} \label{AppSec:DecoyEst}
Here we consider the parameter ($Y_j$ and $e_j$) deviations between the signal states and the decoy states in the case of $N$ discrete phase randomization. Denote the intensity of the signal state to be $\mu$ and decoy state to be $\nu$, $\nu<\mu$. We want to figure out the relationships between $Y_j^\mu$, $e_j^\mu$ and $Y_j^\nu$, $e_j^\nu$, respectively.

We follow the tagged idea for the phase error estimation~\cite {GLLP:2004}. First, we need to evaluate the fidelity between $\ket{\lambda_j^\mu}$ and $\ket{\lambda_j^\nu}$ as defined in Main Text,
\begin{eqnarray} \label{App:SDlambda}
\ket{\lambda_j^\mu} &= &\sum_{l=0}^{\infty} \frac{\alpha^{lN+j}}{\sqrt{(lN+j)!}}\ket{lN+j}\nonumber \\
\ket{\lambda_j^\nu} &=& \sum_{l=0}^{\infty} \frac{\beta^{lN+j}}{\sqrt{(lN+j)!}}\ket{lN+j}
\end{eqnarray}
where $\mu=|\alpha|^2$ and $\nu=|\beta|^2$. We note that these are the states after phase randomization and before qubit encoding. Then the fidelity is given by
\begin{eqnarray} \label{App:SDFidelity}
F(\ket{\lambda_j^\mu},\ket{\lambda_j^\nu}) &=& \frac{|\braket{\lambda_j^\mu}{\lambda_j^\nu}|}{\sqrt{\braket{\lambda_j^\mu}{\lambda_j^\mu} \braket{\lambda_j^\nu}{\lambda_j^\nu}}} \nonumber\\
&=& \frac{|\sum_{l=0}^{\infty} \frac{(\alpha^*\beta)^{lN+j}}{(lN+j)!}|}{\sqrt{\sum_{l=0}^{\infty} \frac{|\alpha|^{2lN+2j}}{(lN+j)!}\sum_{l=0}^{\infty} \frac{|\beta|^{2lN+2j}}{(lN+j)!}}} \nonumber\\
&=& \frac{\sum_{l=0}^{\infty} \frac{(\mu\nu)^{lN/2}}{(lN+j)!}}{\sqrt{\sum_{l=0}^{\infty} \frac{\mu^{lN}}{(lN+j)!}\sum_{l=0}^{\infty} \frac{\nu^{lN}}{(lN+j)!}}}
\end{eqnarray}
In the last equality, we assume $\alpha^*\beta$ is a real number, which can be set when their phases are the same. In experiment, one can think of the scenario where the decoy state intensity modulation is done after phase randomization. When $N\rightarrow\infty$, this fidelity will go to 1 as the photon number channel model. Take the first order approximation when $N$ is large or $\mu$ is small,
\begin{eqnarray} \label{App:SDFApp1st}
F(\ket{\lambda_j^\mu},\ket{\lambda_j^\nu}) &=& \frac{|1+ \frac{(\mu\nu)^{N/2}j!}{(N+j)!}|}{ \left[(1+\frac{\mu^{N}j!}{(N+j)!})(1+\frac{\nu^{N}j!}{(N+j)!}) \right]^{1/2}} +O\left(\left[\frac{\mu^{N}j!}{(N+j)!}\right]^2\right)\\
&=& 1+ \frac{(\mu\nu)^{N/2}j!}{(N+j)!} -\frac12\frac{\mu^{N}j!}{(N+j)!} -\frac12\frac{\nu^{N}j!}{(N+j)!} +O\left(\left[\frac{\mu^{N}j!}{(N+j)!}\right]^2\right)  \nonumber\\
&=& 1- \left[\frac{\mu^N+\nu^{N}}{2} -(\mu\nu)^{N/2}\right]\frac{j!}{(N+j)!} +O\left(\left[\frac{\mu^{N}j!}{(N+j)!}\right]^2\right)\nonumber
\end{eqnarray}
One can show that Eq.~\eqref{App:SDFidelity} is a non-decreasing function with increasing $j$,
\begin{eqnarray} \label{App:SDFmunu}
F(\ket{\lambda_j^\mu},\ket{\lambda_j^\nu}) &\ge& F(\ket{\lambda_0^\mu},\ket{\lambda_0^\nu})\nonumber \\
&=& \frac{\sum_{l=0}^{\infty} \frac{(\mu\nu)^{lN/2}}{(lN)!}}{\sqrt{\sum_{l=0}^{\infty} \frac{\mu^{lN}}{(lN)!}\sum_{l=0}^{\infty} \frac{\nu^{lN}}{(lN)!}}} \nonumber\\
&\equiv & F_{\mu\nu}
\end{eqnarray}

Apply the quantum coin idea from GLLP~\cite {GLLP:2004},
\begin{eqnarray} \label{App:YSDdiff}
\sqrt{Y_j^\mu Y_j^\nu} +\sqrt{(1-Y_j^\mu) (1-Y_j^\nu)} \ge F(\ket{\lambda_j^\mu},\ket{\lambda_j^\nu}) \nonumber \\
\sqrt{e_j^\mu Y_j^\mu e_j^\nu Y_j^\nu} +\sqrt{(1-e_j^\mu Y_j^\mu) (1-e_j^\nu Y_j^\nu)} \ge F(\ket{\lambda_j^\mu},\ket{\lambda_j^\nu})
\end{eqnarray}
Normally $Y_j$ is in the order of channel transmittance $\eta$. One can see that if $F(\ket{\lambda_j^\mu},\ket{\lambda_j^\nu})\le\sqrt{1-\eta}$, the difference can be from $[0,1]$, which would result in zero key rate. On the other hand, if $F=1$, we have $Y_j^\mu=Y_j^\nu$, which is reasonable since the yields of the same states should be the same.

With the calculations presented in Section \ref{AppSub:SolveYSD}, we can solve Eq.~\eqref{App:YSDdiff},
\begin{eqnarray} \label{App:YSDdiff}
|Y_j^\mu-Y_j^\nu| &\le& \sqrt{1-F_{\mu\nu}^2} \nonumber\\
|e_j^\mu Y_j^\mu-e_j^\nu Y_j^\nu| &\le& \sqrt{1-F_{\mu\nu}^2}
\end{eqnarray}
Note that once $N$, $\mu$ and $\nu$ are given, $F_{\mu\nu}$ is given by Eq.~\eqref{App:SDFmunu} and hence the yield and error rate differences are fixed.

\subsection{Bound the parameter difference between signal and decoy state} \label{AppSub:SolveYSD}
To make it simpler, we rewrite Eq.~\eqref{App:YSDdiff} in the following form,
\begin{equation} \label{App:YSDdiffEq}
\sqrt{a b} +\sqrt{(1-a) (1-b)} \ge F \\
\end{equation}
where $a,b\in[0,1]$. Let $a=\sin^2 x$ and $b=\sin^2 y$, where $x,y\in[0,\pi/2]$, then
\begin{eqnarray} \label{App:Ysincos}
F &\le& \sin x \sin y +\cos x \cos y \nonumber\\
&=& \cos(x-y)
\end{eqnarray}
Thus,
\begin{equation} \label{App:Ysolve}
|x-y| \le \arccos F
\end{equation}
Since $F$ is close 1, $|x-y|$ is close to 0. That is, $a$ and $b$ are close to each other,
\begin{eqnarray} \label{App:Yab}
|a-b| &=& |\sin^2x-\sin^2y| \nonumber\\
&=& |\sin(x+y)\sin(x-y)| \nonumber\\
&\le& \sin(\arccos F) \\
&=& \sqrt{1-F^2}  \nonumber
\end{eqnarray}

\section{Simulation}
\label{AppSec:Sim}
In this section, we calculate the key rates of both decoy and non-decoy methods derived in the Main text.
We use typical experimental parameters~\cite {GYS:122km:2004} which are $e_d=0.033$, $\eta=10^{-\alpha L/10}\eta_{Bob}$ where
$\alpha=0.2~\mathrm{dB/km}$,  $\eta_{Bob}=0.045$, $Y_0=1.7\times 10^{-6}$
and assumed an error-correction inefficiency $f(e)=1.16$. Here $e_d$ is the intrinsic error rate of Bob's detectors. For each value of the distance, the signal strength $\mu$ has been chosen to optimize the rate. In the simulation model, $Q_\mu=Y_0+1-e^{-\eta\mu}$.

\subsection{Non-decoy}

1. First we calculate $P_j= \sum_{l=0}^{\infty}\frac{\mu^{lN+j}e^{-\mu}}{(lN+j)!}$.

\noindent 2. Then we calculate $F_j(\rho_x,\rho_y) \ge \left| \frac{\sum_{l=0}^{\infty} \frac{\mu^{lN+j}}{(lN+j)!} 2^{-\frac{lN+j}{2}} \left(\cos\frac{lN+j}{4}\pi+\sin\frac{lN+j}{4}\pi\right) }{\sum_{l=0}^{\infty}\frac{\mu^{lN+j}}{(lN+j)!}} \right|$.

\noindent 3. For $(e_0, e_1, Y_0, Y_1)$ in the domain defined by
\begin{eqnarray*}
P_0 Y_0 +P_1 Y_1 &\ge& Q_\mu-\sum_{j=2}^{N-1} P_j, \nonumber \\
e_0P_0 Y_0 +e_1P_1 Y_1 &\le& E_\mu Q_\mu
\end{eqnarray*}
according to the Main text where the notations are defined in the Main text, we calculate $\Delta_j$ and $e_j^p$ according to Eq.~\eqref{Discrete:bitphase} and \eqref{Discrete:bias}.

\noindent 4. Substitute the above quantities into $\min_{0\le Y_j,e_j^b\le 1} \{P_0 Y_0[1-H(e_0^p)] +P_1 Y_1[1-H(e_1^p)]\}$  and numerically optimize $(e_0, e_1, Y_0, Y_1)$ for the minimum.

\noindent 5. Calculate the key rate $R=\min_{0\le Y_j,e_j^b\le 1} \{P_0 Y_0[1-H(e_0^p)] +P_1 Y_1[1-H(e_1^p)]\}-I_{ec}$.

The signal intensity $\mu$ is numerically optimized to maximize the key rate. A typical value of $\mu$ ranges from 0.001 to 0.02. When the number of phases $N$ is large, $\mu$ is approximately the decay rate $\eta$.

\subsection{Decoy}

1. First we calculate $P_j= \sum_{l=0}^{\infty}\frac{\mu^{lN+j}e^{-\mu}}{(lN+j)!}$.

\noindent 2. Then we calculate $F_j(\rho_x,\rho_y) \ge \left| \frac{\sum_{l=0}^{\infty} \frac{\mu^{lN+j}}{(lN+j)!} 2^{-\frac{lN+j}{2}} \left(\cos\frac{lN+j}{4}\pi+\sin\frac{lN+j}{4}\pi\right) }{\sum_{l=0}^{\infty}\frac{\mu^{lN+j}}{(lN+j)!}} \right|$.

\noindent 3. Next we calculate $F_{\mu\nu} =  1- O\left(\frac{\mu^{N}}{N!}\right)$.

\noindent 4. For $(e_0, e_1, Y_0, Y_1)$ in the domain defined by
 \begin{eqnarray*}
|Y_j^\mu-Y_j^\nu| \le \sqrt{1-F_{\mu\nu}^2} \\
|e_j^\mu Y_j^\mu-e_j^\nu Y_j^\nu| \le \sqrt{1-F_{\mu\nu}^2} \\
Q_\mu = \sum_{j=0}^{N-1} P_j^\mu Y_j^\mu \\
E_\mu Q_\mu = \sum_{j=0}^{N-1} e_j^\mu P_j^\mu Y_j^\mu,
\end{eqnarray*}  we calculate $\Delta_j$ and $e_j^p$ according to Eq.~\eqref{Discrete:bitphase} and \eqref{Discrete:bias}.

\noindent 5. Substitute the above quantities into $\min_{0\le Y_j,e_j^b\le1} \{P_0 Y_0[1-H(e_0^p)] +P_1 Y_1[1-H(e_1^p)]\}$ and numerically optimize $(e_0, e_1, Y_0, Y_1)$ for the minimum.

\noindent 6. Calculate the key rate $R=\min_{0\le Y_j,e_j^b\le 1} \{P_0 Y_0[1-H(e_0^p)] +P_1 Y_1[1-H(e_1^p)]\}-I_{ec}$.

The decoy and signal intensities $\mu$ and $\nu$ are numerically optimized to maximize the key rate. A typical value of $\mu$ is 0.5. One weak decoy state with a typical mean photon number of $\nu=0.001$ and one vacuum state are used.

\section*{References}

 \bibliographystyle{iopart-num}

\bibliography{BibliDiscrete}


\end{document}